 \documentclass[final,5p,times,twocolumn]{elsarticle}

\usepackage{amssymb}
\usepackage{amsmath}
\usepackage[utf8]{inputenc}

\journal{Physica E}

\begin{document}

\begin{frontmatter}



\title{Floquet Majorana Fermions in superconducting quantum dots}


\author[icmm,gar]{M\'onica Benito}
\ead{m.benito@csic.es}
\author[icmm]{Gloria Platero}
\address[icmm]{Instituto de Ciencia de Materiales, CSIC, Cantoblanco, Madrid E-28049,
Spain\\}
\address[gar]{Max-Planck Institut f\"{u}r Quantenoptik, Hans-Kopfermann-Str. 1, D-85748 Garching, Germany\\}

\begin{abstract}
We consider different configurations of ac driven quantum dots coupled
to superconductor leads where Majorana fermions can exist as collective
quasiparticles. The main goal is to tune the existence, localization
and properties of these zero energy quasiparticles by means of periodically
driven external gates. In particular, we analyze the relevance of
the system and driving symmetry. We predict the existence of different
sweet spots with Floquet Majorana fermions in configurations where
they are not present in the undriven system.
\end{abstract}


\end{frontmatter}


\section{Introduction}
\label{intro}
There are condensed matter systems which can hold collective quasiparticles that
are their own antiparticles, therefore satisfying the Majorana condition \cite{Wilczek2009,Alicea2012,Beenakker2013}. These quasiparticles are termed Majorana 
Fermions (MFs) and follow non-abelian statistics. Detection of MFs in solid state systems have been  recently experimentally  proposed \cite{Mourik2012,Rokhinson2012,Nadj-perge2014}.
Recently, the interest in encoding a qubit in these kind of excitations
has grown due to the possibility to be non-local, a property which has a great
potential in quantum computation due to the robustness of the qubit
against local perturbations \cite{Nayak2008}. Furthermore, how to  tune MFs in condensed matter systems is one of the main purposes of research in the emergent field of topological quantum computation.\\
In the last years, different
works have shown how the application of ac fields enriches the properties
of these quasiparticles and facilitate their tunability. For
instance, it is possible to generate Floquet Majorana fermions (FMFs)
as steady-states of non-equilibrium systems which present  interesting
properties for quantum computation: non-locality and non-abelian statistics \cite{Liu2013,Jiang2011}.

In every system with particle-hole symmetry, the quasiparticles come
in pairs $\gamma_{-E}^{\dagger}=\gamma_{E}$, therefore they can hold
MFs as long as the energy can be tuned to zero.
One of the simplest and most tunable system with particle hole symmetry is
a double quantum dot (QD) connected via an s-wave superconductor \cite{Leijnse2012}.
It is well known that the proximity effect induces Cooper pairs correlations
across the dots \cite{Martin-Rodero2011,DeFranceschi2010} generating
effectively superconductivity \cite{Weiss2014}. 
Interestingly, Fractional Josephson effect, a signature of the presence of MFs \cite{Rokhinson2012,PhysRevB.86.140503,Jacquod2013},  in a quadruple
quantum dot in the presence of an s-wave superconductor has been predicted by Markus  B\"{u}ttiker and coworkers \cite{Sothmann2013}.

The advantage that configurations of a few QDs connected to  s-wave superconductors present, in order to generate and detect MFs, in comparison with nano-wires \cite{Leijnse2012a,Lutchyn2010,Oreg2010} 
or long QD chains \cite{Kitaev2001,2012NatCo...3E.964S,Fulga2013} proposals
is their great tunability, while in the latter the MFs have topological protection.

In this paper we analyze two different configurations of QDs in proximity
to superconducting leads such that Cooper pair correlations are induced
between the neighboring dots as long as the coherence length is larger
than the distance between them. We include periodically driven gates and search
for the conditions for appearance of FMFs. The paper is organized
as follows: In section \ref{undriven} we present the model, in section \ref{Floquet}
we discuss the generation of FMFs in a double and a triple superconducting
QD. Finally, we present our conclusions in section \ref{conclu}. 

\section{Undriven system}
\label{undriven}

Systems of QDs coupled to s-wave superconductors have been a subject
of study \cite{Martin-Rodero2011,Weiss2014,Sothmann2013} 
\begin{figure}
\centering
\includegraphics[width=0.85\columnwidth]{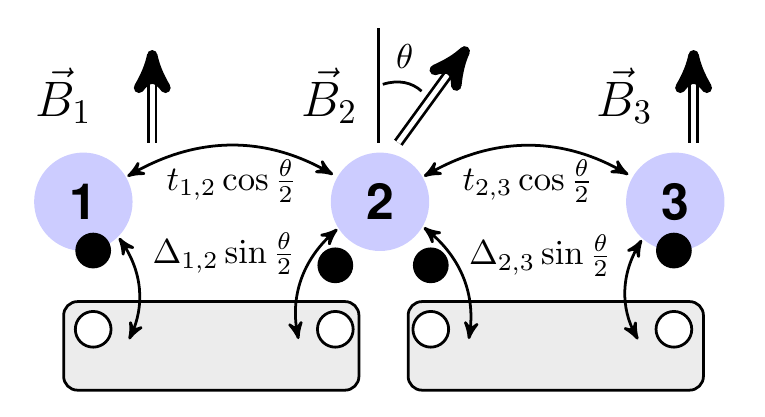}

\caption{\label{fig:Scheme-of-the}Scheme of three QDs coupled by tunnel and coupled to a superconductor. 
The existence of Cooper pairs generates correlations of the
type $d_{i,\sigma}d_{i+1,\bar{\sigma}}$ in the effective Hamiltonian
for the QDs. The applied magnetic fields and their directions
are also shown in the picture. The angle $\theta$ controls the ratio
$\Delta_{i,i+1}/t_{i,i+1}$ (see text below).}

\end{figure}
 because the proximity effect induces Cooper pair correlations that
can be easily detected due to the low number of degrees of freedom
in QDs. In a system where neighboring QDs are coupled through
superconducting reservoirs as in Fig.\ref{fig:Scheme-of-the}, in the limit of large superconducting gap 
the superconductors can be traced out and an effective
Hamiltonian for the dots is obtained \cite{Wright2013,Eldridge2010}
\begin{eqnarray}
H & = & \sum_{i,\sigma}\mu_{i,\sigma}d_{i,\sigma}^{\dagger}d_{i,\sigma}\label{eq:H1}\\
 &  & +\sum_{i,\sigma}\left(t_{i,i+1}d_{i,\sigma}^{\dagger}d_{i+1,\sigma}+\Delta_{i,i+1}d_{i,\sigma}d_{i+1,\bar{\sigma}}+h.c.\right)\ ,\nonumber 
\end{eqnarray}
which already contains effective superconductivity between neighboring
dots. The fermionic operator $d_{i,\sigma}$ represents the annihilation
of an electron in the $i$-QD with spin $\sigma$. The symbol $\bar{\sigma}$
means the opposite spin to $\sigma$, which can be $\sigma=\uparrow,\downarrow$.
$\mu_{i}$ is the onsite energy in  $i$-QD, the parameter $t_{i,i+1}$
is the effective tunneling probability from dot $i$ to dot $i+1$
through the superconductor by virtual occupation of the above gap
excitations and $\Delta_{i,i+1}$ is the effective superconducting
amplitude due to the superconductor connecting  the $i$ and $i+1$ dots. If
a large magnetic field is applied to the dots only one spin comes
into play. However, the magnetic fields have to be non-collinear in
order to have s-wave type Cooper pair correlations (see Fig.\ref{fig:Scheme-of-the})
\cite{Leijnse2012}. In this configuration, it is more natural to
work in the basis of the quantization axes given by the magnetic field
in each dot. For that purpose, we
have to perform the rotation
\begin{equation}
d_{2,\sigma}\rightarrow\cos\frac{\theta}{2}d_{2,\sigma}+\sigma\sin\frac{\theta}{2}d_{2,\bar{\sigma}}\label{eq:transf}
\end{equation}
as the magnetic field in the central QD forms an angle $\theta$
with the magnetic fields in the left and right QDs (see Fig.\ref{fig:Scheme-of-the}).
The low-frequency hamiltonian will be given by eq.(\ref{eq:H1})
by neglecting the contribution from the high-energy spin direction
in each dot (keeping $\sigma=\downarrow$): 
\begin{eqnarray}
H & = & \sum_{i}\mu_{i}d_{i}^{\dagger}d_{i}\label{eq:H1spin}\\
 &  & +\sum_{i}\left\{ t'_{i,i+1}d_{i}^{\dagger}d_{i+1}+\Delta'_{i,i+1}d_{i}d_{i+1}+h.c.\right\}\ , \nonumber 
\end{eqnarray}
where $d_{i}\equiv d_{i,\downarrow}$, $t'_{i,i+1}\equiv t_{i,i+1}\cos\frac{\theta}{2}$
and $\Delta'_{i,i+1}\equiv \Delta_{i,i+1}\sin\frac{\theta}{2}$. Therefore 
the normal and superconducting tunneling amplitudes are renormalized and their renormalization depends on the angle between the magnetic field directions. This dependence introduces a simple way to tune externally the coupling parameters of the system \cite{Leijnse2012}.\\
 In order to obtain the excitation
spectrum of the system  the Hamiltonian is written in the Nambu
basis $\Psi=\begin{pmatrix}d_{1}, & d_{1}^{\dagger}, & ..., & ..., & d_{N}, & d_{N}^{\dagger}\end{pmatrix}$
as:
\begin{equation}
H=\frac{1}{2}\Psi^{\dagger}h\Psi+\frac{1}{2}\sum_{i}\mu_{i}\ .\label{eq:Hnambu}
\end{equation}
For a triple QD $h$ reads
\begin{equation}
h=\begin{pmatrix}\mu_{1} & 0 & t'_{1,2} & -\Delta'_{1,2} & 0 & 0\\
0 & -\mu_{1} & \Delta'_{1,2} & -t'_{1,2} & 0 & 0\\
t'_{1,2} & \Delta'_{1,2} & \mu_{2} & 0 & t'_{2,3} & -\Delta'_{2,3}\\
-\Delta'_{1,2} & -t'_{1,2} & 0 & -\mu_{2} & \Delta'_{2,3} & -t'_{2,3}\\
0 & 0 & t'_{2,3} & \Delta'_{2,3} & \mu_{3} & 0\\
0 & 0 & -\Delta'_{2,3} & -t'_{2,3} & 0 & -\mu_{3}
\end{pmatrix}\ ,\label{eq:Hmatrix}
\end{equation}
The eigensystem of $h$ ($h{\bf v}_{i}=\lambda_{i}{\bf v}_{i}$) determines the quasiparticles, given by $\gamma_{i}={\bf v}_{i}\cdot\Psi$. A
zero-energy solution, $\lambda_{i}=0$, implies the presence of a pair of Majorana quasiparticles.

In the case of a double QD one can choose an angle such that $\Delta'_{1,2}=\pm t'_{1,2}$
and if $\mu_{1}=0$, there are two MFs given by
\begin{eqnarray}
\gamma_{1} & = & \frac{1}{\sqrt{2}}\left(d_{1}\mp d_{1}^{\dagger}\right)\ ,\label{eq:states}\\
\gamma_{2} & = & \frac{1}{\sqrt{2}\sqrt{1+\delta^{2}}}\left\{ \left(d_{2}\pm d_{2}^{\dagger}\right)-\delta\left(d_{1}\pm d_{1}^{\dagger}\right)\right\}\ , \nonumber 
\end{eqnarray}
where $\delta=\frac{\mu_{2}}{2t'_{1,2}}$. Only in the case where
$\mu_{2}=0$ the MFs are spatially separated \cite{Leijnse2012}.
In the case of a triple QD, assuming $\Delta'_{i,i+1}=\pm t'_{i,i+1}$ and $\mu_{1}=0$,
there are  two MFs  given by
\begin{eqnarray}
\gamma_{1} & = & \frac{1}{\sqrt{2}}\left(d_{1}\mp d_{1}^{\dagger}\right)\ ,\label{eq:states3}\\
\gamma_{2} & = & \frac{\left(d_{3}\pm d_{3}^{\dagger}\right)-\alpha\left(d_{2}\pm d_{2}^{\dagger}\right)+\beta\left(d_{1}\pm d_{1}^{\dagger}\right)}{\sqrt{2}\sqrt{1+\alpha^{2}+\beta^{2}}}\ ,\nonumber 
\end{eqnarray}
where $\alpha=\frac{\mu_{3}}{2t'_{2,3}}$ and $\beta=\frac{\mu_{2}\mu_{3}}{4t'_{12}t'_{23}}$.
In the case where  $\mu_{2}$ or $\mu_{3}$  are zero the MFs
are spatially separated \cite{Deng2015}. Interestingly, the manipulation
of the onsite-energies allows to change the localization of the MFs,
which would be  relevant for their detection in  transport \cite{Leijnse2012}.

\section{Floquet Majorana fermions\label{Floquet}}

In the following, we will apply external ac fields in order to change
periodically the onsite energies of the QDs and in this way obtain
FMFs as steady-state solutions of the non-equilibrium problem.\\
 For
every system described by a time-periodic Hamiltonian 
a set of solutions exists, called Floquet states, which have the form $|\psi_{n}(t)\rangle=e^{-i\epsilon_{n}t}|u_{n}(t)\rangle$,
where $|u_{n}(t)\rangle$ are time periodic functions called Floquet
modes and $\epsilon_{n}$ are the so called quasienergies \cite{Platero2004,Sambe1973,Shirley1965}.
As the quasienergies are only defined modulo $\Omega$, where $\Omega=\frac{2\pi}{T}$
and $T$ is the period of the Hamiltonian, a system with particle-hole
symmetry (with excitations in pairs $\gamma_{-\epsilon}^{\dagger}=\gamma_{\epsilon}$) will hold FMFs if $\epsilon=0,\pm\Omega/2$. If the frequency
is large enough, it is a good approximation to consider the time-averaged
Hamiltonian to describe the dynamics. For lower frequencies, where multi photon processes are relevant , the dynamics becomes more involved 
but there is also a way to find  an effective  time-independent hamiltonian  which includes as many photon processes as necessary \cite{A.EckardtandE.Anisimovas,Goldman2014}.

The motivation to consider periodically driven quantum
systems is the fact that their time-evolution is governed by an effective time-independent Hamiltonian, whose properties can be engineered according to the particular purposes. This method, called Floquet engineering,
has been employed to achieve dynamic localization \cite{Grossmann1991,Creffield2004,PhysRevB.84.121310},
photon-assisted tunneling \cite{Platero2004,Gallego-marcos2014} or nobel
topological band structures \cite{Inoue2010,Lindner2011,Gomez-Leon2013,DelPlace2013,Grushin2014,Gomez-Leon2014,Benito2014}. \\
The application of degenerate perturbation theory in the extended
Floquet Hilbert space provides a high-frequency expansion (in powers
of $\frac{1}{\Omega}$) for this effective Hamiltonian, such as $H_{F}=\sum_{\nu=0}^{\infty}H_{F}^{\nu}$
 \cite{A.EckardtandE.Anisimovas}. With the definition of  the Fourier
components of the time-periodic Hamiltonian
\begin{equation}
H_{m}=\frac{1}{T}\int_{0}^{T}dte^{-im\Omega t}H(t)\ ,\label{eq:fourier}
\end{equation}
the leading orders of the expansion for the effective Hamiltonian
are{\small{}
\begin{eqnarray}
H_{F}^{0} & = & H_{0}\ ,\nonumber \\
H_{F}^{1} & = & \sum_{m=1}^{\infty}\frac{\left[H_{m},H_{-m}\right]}{m\Omega}\ ,\label{eq:corrections}\\
H_{F}^{2} & = & \sum_{m\neq0}\left(\frac{\left[H_{-m},\left[H_{0},H_{m}\right]\right]}{2\left(m\Omega\right)^{2}}+\sum_{m'\neq0,m}\frac{\left[H_{-m'},\left[H_{m'-m},H_{m}\right]\right]}{3mm'\left(\Omega\right)^{2}}\right)\ .\nonumber 
\end{eqnarray}}
These terms will be considered below in order to obtain FMFs in two different configurations of driven quantum dots: DQDs and TQDs.\\
The time periodic perturbation applied to the i-QD is:
\begin{equation}
V(t)=\sum_{i}A_{i}\cos\left(\Omega t+\varphi_{i}\right)d_{i}^{\dagger}d_{i}\ .\label{eq:perturbation}
\end{equation}
In order to study the effect of an external driving at high frequency,
it is convenient to move to the interaction picture which transfers the time-dependence
to the tunneling terms by means of the unitary transformation: $U(t)=\exp\left\{ -i\int_{0}^{t}V(t')dt'\right\} $. Only the non-diagonal elements change under the transformation
depending on whether they commute or not with the time-periodic term:
\begin{eqnarray}
\left[d_{i}^{\dagger}d_{i+1},V(t)\right] & = & d_{i}^{\dagger}d_{i+1}\left(A_{i+1}\cos\left(\Omega t+\varphi_{i+1}\right)\right.\nonumber \\
 &  & \left.-A_{i}\cos\left(\Omega t+\varphi_{i}\right)\right)\ ,\nonumber \\
\left[d_{i}d_{i+1},V(t)\right] & = & d_{i}d_{i+1}\left(A_{i+1}\cos\left(\Omega t+\varphi_{i+1}\right)\right.\nonumber \\
 &  & \left.+A_{i}\cos\left(\Omega t+\varphi_{i}\right)\right)\ .\label{eq:conmutation}
\end{eqnarray}
Therefore, the renormalization of the tunneling and the superconducting
pairing depends on the symmetry of the driving i.e., on the intensities applied in the different dots and on the phase difference of the ac gate voltages between the different dots. As
an example of this, in the case of two QDs if an ac gate potential
is applied to each of them with the same amplitude ($A_{1}=A_{2}$)
and frequency, the tunneling term does not change if the phases are
equal but it does if the phase difference is $\pi$ and the opposite happens
for the superconducting pairing (see eq.\ref{eq:conmutation}) \cite{Li,Li2014}. 

\subsection{Superconducting double QD:}

In the present work, we are interested in a configuration such that
both the tunnel and the superconducting amplitudes are equaly renormalized
by the ac voltages. By inspection of eq.(\ref{eq:conmutation}),
one can see that this corresponds to driving one of the gates periodically,
it means $A_{1}=A_{0}$ and $A_{2}=0$. In this case, the Fourier
components of the time-dependent Hamiltonian are
\begin{equation}
H_{m}=\begin{pmatrix}\mu_{1}\delta_{n,0} & 0 & t_{m} & -\Delta_{m}\\
0 & -\mu_{1}\delta_{n,0} & \Delta_{-m} & -t_{-m}\\
t_{-m} & \Delta_{m} & \mu_{2}\delta_{n,0} & 0\\
-\Delta_{-m} & -t_{m} & 0 & -\mu_{2}\delta_{n,0}
\end{pmatrix}\ ,\label{eq:bessel2dots}
\end{equation}
where $t_{m}\equiv t'_{1,2}{\cal J}_{m}\left(\frac{A_{0}}{\Omega}\right)$,  $\Delta_{m}\equiv \Delta'_{1,2}{\cal J}_{m}\left(\frac{A_{0}}{\Omega}\right)$
and ${\cal J}_{m}$ is the $m$-Bessel function of first kind. The
zeroth order effective Hamiltonian only predicts spatially separated
FMFs if $\mu_{1,2}=0$ and $\Delta'_{1,2}=\pm t'_{1,2}$ (see eq.\ref{eq:states}). However, the following order corrections allow
to generate new sweet spots for FMFs. The first order correction is
zero and the effect of the second one is the renormalization of $t'_{1,2}$
and $\Delta'_{1,2}$ to some effective values given by:
\begin{eqnarray}
t_{\text{eff}} & = & t'_{1,2}{\cal J}_{0}\left(\frac{A_{0}}{\Omega}\right)-\frac{4 {t'}_{1,2}^{2}}{\Omega^{2}}f\left(\frac{A_{0}}{\Omega}\right)\ ,\label{eq:teff2dots}\\
\Delta_{\text{eff}} & = & \Delta'_{1,2}{\cal J}_{0}\left(\frac{A_{0}}{\Omega}\right)-\frac{4{\Delta'}_{1,2}^{2}}{\Omega^{2}}f\left(\frac{A_{0}}{\Omega}\right)\ ,\label{eq:deltaeff2dots}
\end{eqnarray}
where $f\left(\frac{A_{0}}{\Omega}\right)$ is a function of all the
Bessel functions. Considering only two sidebands (${\cal J}_{n,-n}\left(\frac{A_{0}}{\Omega}\right)=0$
for $n>2$) its analytical expression becomes:
\begin{eqnarray}
 f\left(\frac{A_{0}}{\Omega}\right)={\cal J}_{1}^{2}\left(\frac{A_{0}}{\Omega}\right)\left({\cal J}_{0}\left(\frac{A_{0}}{\Omega}\right)+{\cal J}_{2}\left(\frac{A_{0}}{\Omega}\right)\right)\ .
 \end{eqnarray}
 As the ratio between the intensity and the frequency of the ac field increases more terms contribute to $f\left(\frac{A_{0}}{\Omega}\right)$ . \\
 The key point in the previous discussion is that the renormalization of  $t'_{1,2}$
and $\Delta'_{1,2}$  by the ac field makes it possible to choose the driving
amplitude such that $\Delta_{\text{eff}}=\pm t_{\text{eff}}$ even
when $\Delta'_{1,2}\neq t'_{1,2}$. 
\begin{figure}
\centering
\includegraphics[width=1\columnwidth]{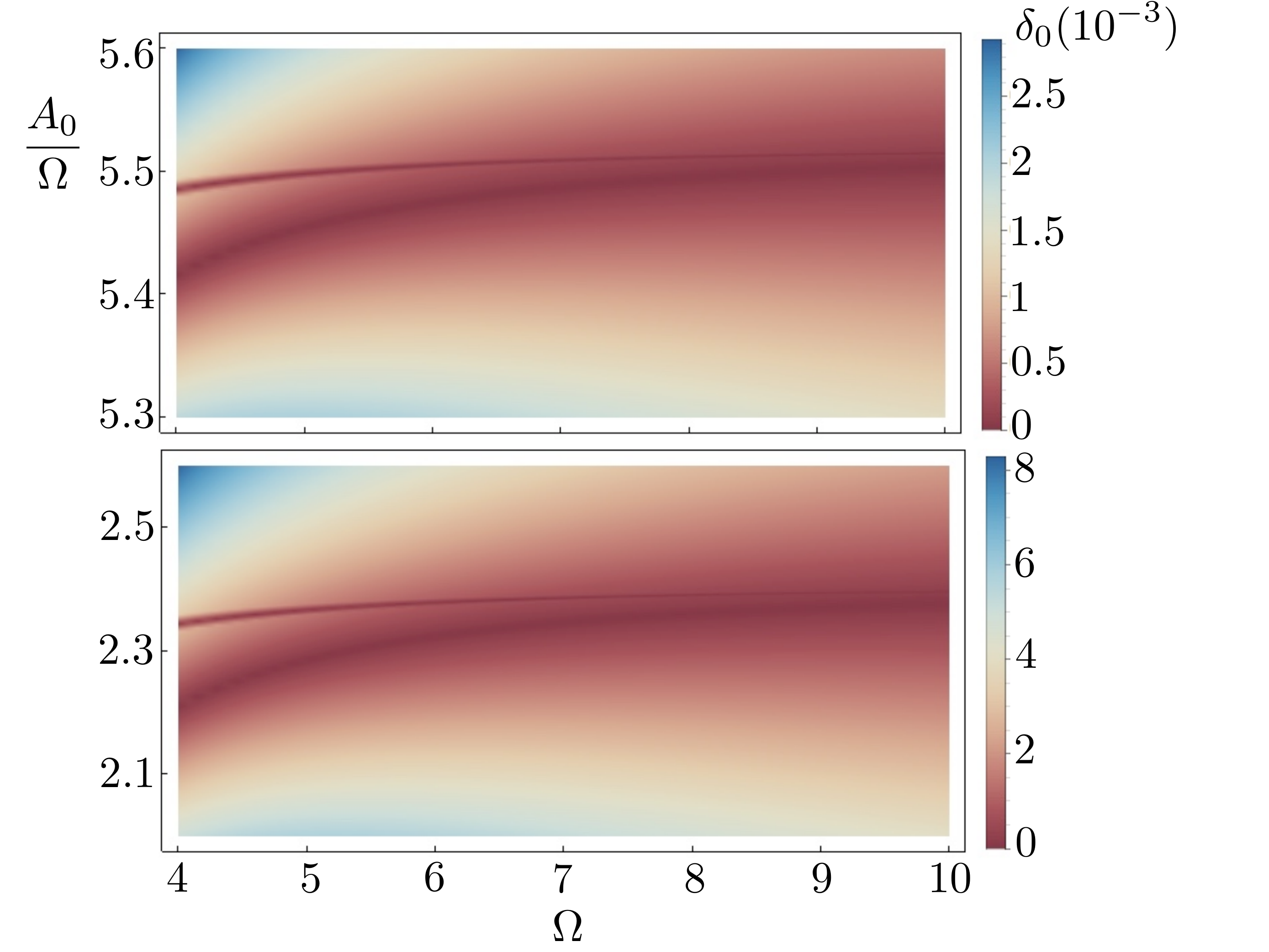}

\protect\caption{\label{fig:2dots-splitting-1}Quasienergy gap $\delta_0$ for a superconducting
double QD as a function of the amplitude and frequency of the driving.
The dark regions corresponds to closed gap, i.e., zero quasienergy.
The plot shows that the 4-fold degeneracy at high frequency at the
zeros of the Bessel function ${\cal J}_{0}\left(\frac{A_{0}}{\Omega}\right)$
splits into two different sweet spots with FMFs as the frequency decreases.
The bottom plot shows the region around the first zero and the upper
plot around the second zero. Parameters: $\mu_{1}=\mu_{2}=0$, $\Delta'_{1,2}=1$,
$t'_{1,2}=0.8$, $\varphi_{3}-\varphi_{1}=0$. All the energies are in units of $\Delta'_{1,2}$, which is set to 1.}
\end{figure}
 This is exactly what we observe in the
quasienergy spectrum (see Fig.\ref{fig:2dots-splitting-1}). In this
calculation, the on-site energies $\mu_{1}$ and $\mu_{2}$ are set
to zero and the static normal and superconducting  tunnelings are different, i.e., $\Delta'_{1,2}\neq t'_{1,2}$. At high frequencies all the quasienergies are zero  at the zeros of the function ${\cal J}_{0}\left(\frac{A_{0}}{\Omega}\right)$
(approximately $\frac{A_{0}}{\Omega}=2.40,5.52,8.65...$) and there
are no FMFs. As the second order correction becomes important, i.e., as the frequency decreases,
two
different driving amplitudes allow for the condition required to the existence of FMFs: the one for which $\Delta_{\text{eff}}=t_{\text{eff}}$
and the one for which $\Delta_{\text{eff}}=-t_{\text{eff}}$. This
is why at lower frequencies  there are two quasienergy gap closings
around each zero of the Bessel function, i.e., two different sweet spots (the bottom panel of Fig.\ref{fig:2dots-splitting-1} shows the gap around the first zero,
$\sim2.40$ and the upper panel around the second one $\sim5.52$).
In the following we generalize this method for generation of FMFs
to a largest system, i.e., to an array of three QDs. 

\subsection{Superconducting triple QD:}

\begin{figure}
\centering
\includegraphics[width=1\columnwidth]{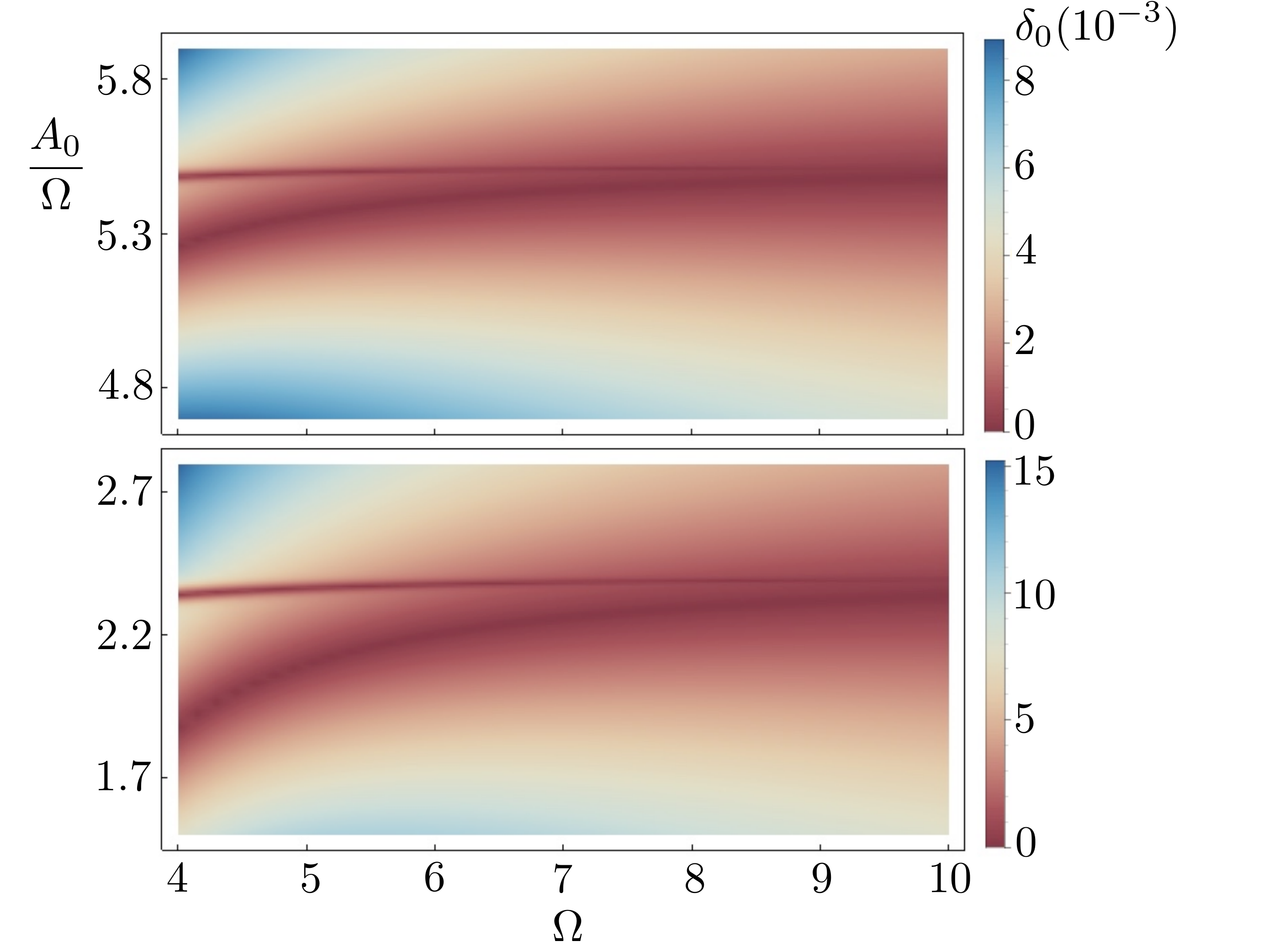}

\protect\caption{\label{fig:splitting}Quasienergy gap  $\delta_0$ for a superconducting triple
QD as a function of the amplitude and frequency of the driving. The
dark regions corresponds to closed gap, it means, zero quasienergy.
The plot shows that the 4-fold degeneracy at high frequency at the
zero of the Bessel function ${\cal J}_{0}\left(\frac{A_{0}}{\Omega}\right)$
splits into two different sweet spots with FMFs as the frequency decreases.
The bottom plot shows the region around the first zero and the upper
plot around the second zero. Parameters: $\mu_{1}=\mu_{2}=0$, $\mu_{3}=1.5$,
$\Delta=1$, $t=0.8$, $\varphi_{3}-\varphi_{1}=0$. All the energies are in units of $\Delta$, which is set to 1.}
\end{figure}

\begin{figure}
\centering
\includegraphics[width=0.75\columnwidth]{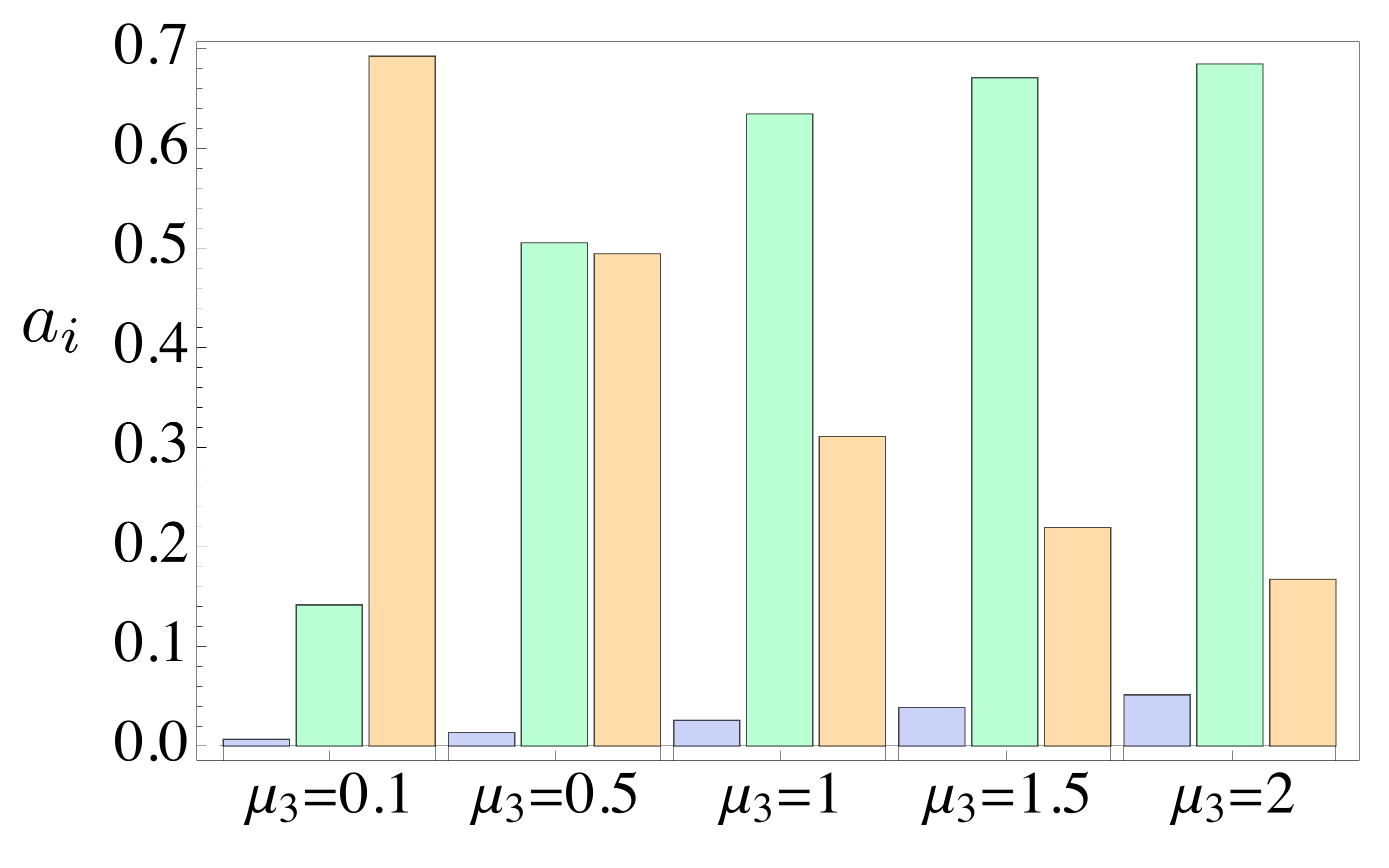}

\protect\caption{\label{fig:loc}Spatial location of the second FMF appearing when
$t_{\text{eff}}=\Delta_{\text{eff}}$. The blue, green and orange
bars are respectively $a_{1}$, $a_{2}$ and $a_{3}$ (see eq.(\ref{eq:FMFs})).
Parameters: $\mu_{1}=\mu_{2}=0$, $\Delta=1$, $t=0.8$, $\Omega=4$,
$\varphi_{3}-\varphi_{1}=0$. All the energies are in units of $\Delta$, which is set to 1.}
\end{figure}

Analogously to the case of the double QD, we use the driving fields
such that all the non-diagonal terms of the Hamiltonian are renormalized
in the same way by the ac field at high-frequency. That implies driving
the left and right dots with ac gate voltages such that $A_{1}=A_{3}=A_{0}$
and $A_{2}=0$. Let us choose for simplicity $t\equiv t'_{1,2}=t'_{2,3}$
and $\Delta\equiv \Delta'_{1,2}=\Delta'_{2,3}$. We are going to analyze
the presence of FMF as a function of the different parameters of the
present setup, in particular of the phase difference between the ac
voltages. With the driving fields in phase $\varphi\equiv \varphi_{3}-\varphi_{1}=0$,
the Fourier components of the time-dependent Hamiltonian are
\begin{equation}
H_{m}=\begin{pmatrix}\mu_{1}\delta_{n,0} & 0 & t_{m} & -\Delta_{m} & 0 & 0\\
0 & -\mu_{1}\delta_{n,0} & \Delta_{-m} & -t_{-m} & 0 & 0\\
t_{-m} & \Delta_{m} & \mu_{2}\delta_{n,0} & 0 & t_{-m} & -\Delta_{m}\\
-\Delta_{-m} & -t_{m} & 0 & -\mu_{2}\delta_{n,0} & \Delta_{-m} & -t_{m}\\
0 & 0 & t_{m} & \Delta_{m} & \mu_{3}\delta_{n,0} & 0\\
0 & 0 & -\Delta_{-m} & -t_{-m} & 0 & -\mu_{3}\delta_{n,0}
\end{pmatrix}\label{eq:bessel}
\end{equation}
where $t_{m}\equiv t{\cal J}_{m}\left(\frac{A_{0}}{\Omega}\right)$
and $\Delta_{m}\equiv \Delta{\cal J}_{m}\left(\frac{A_{0}}{\Omega}\right)$. Due to the driving symmetry, if we keep only the zero
order term of the expansion for the effective Hamiltonian all the
non-diagonal terms vanish at the zeros of ${\cal J}_{0}\left(\frac{A_{0}}{\Omega}\right)$
so there is no effective tunneling or superconducting pairing and
the quasienergies are $\pm\mu_{i}$ for $i=1,2,3$. In the following,
we show how the higher order corrections to this high-frequency approximation generate
FMFs around these zeros. We will focus on the case $\mu_{1}=\mu_{2}=0$,
$\mu_{3}\neq0$ and $\Delta'_{i}\neq\pm t'_{i}$ such that there are not MFs in the static case.
In Fig.\ref{fig:splitting},
we plot the gap of the quasienergy spectrum as a function of the amplitude
and the frequency of the driving. In the limit of high-frequency the
effective tunneling and superconducting pairing are zero so there
is a four-fold degeneracy at $\epsilon=0$ and there are no MFs.
At lower frequencies, these zero-quasienergy pairs appear at
different amplitudes in which the Majorana condition is satisfied,
two different sweet spots. This is due to the second order correction
to the effective Hamiltonian. The largest effect of this term is a
correction of the tunneling amplitudes, which become:
\begin{eqnarray}
t_{\text{eff}} & = & t{\cal J}_{0}\left(\frac{A_{0}}{\Omega}\right)-\frac{4t\left(2t^{2}-\Delta^{2}\right)}{\Omega^{2}}f\left(\frac{A_{0}}{\Omega}\right)\ ,\label{eq:teff3dots}\\
\Delta_{\text{eff}} & = & \Delta{\cal J}_{0}\left(\frac{A_{0}}{\Omega}\right)-\frac{4\Delta\left(2\Delta^{2}-t^{2}\right)}{\Omega^{2}}f\left(\frac{A_{0}}{\Omega}\right)\ .\label{eq:deltaeff3dots}
\end{eqnarray}
One difference with the double QD system is that in this
case a small effective tunneling between dots 1 and 3 appears due
to virtual processes. The expression for this long-range tunneling is
\begin{equation}
\tau_{1,3}=\frac{\mu_{3}\left(\Delta^{2}-t^{2}\right)}{\Omega^{2}}\sum_{m=1}^{\infty}\frac{{\cal J}_{m}\left(\frac{A_{0}}{\Omega}\right)^{2}}{m^{2}}\ .\label{eq:tau13}
\end{equation}
\begin{figure}
\centering
\includegraphics[width=0.75\columnwidth]{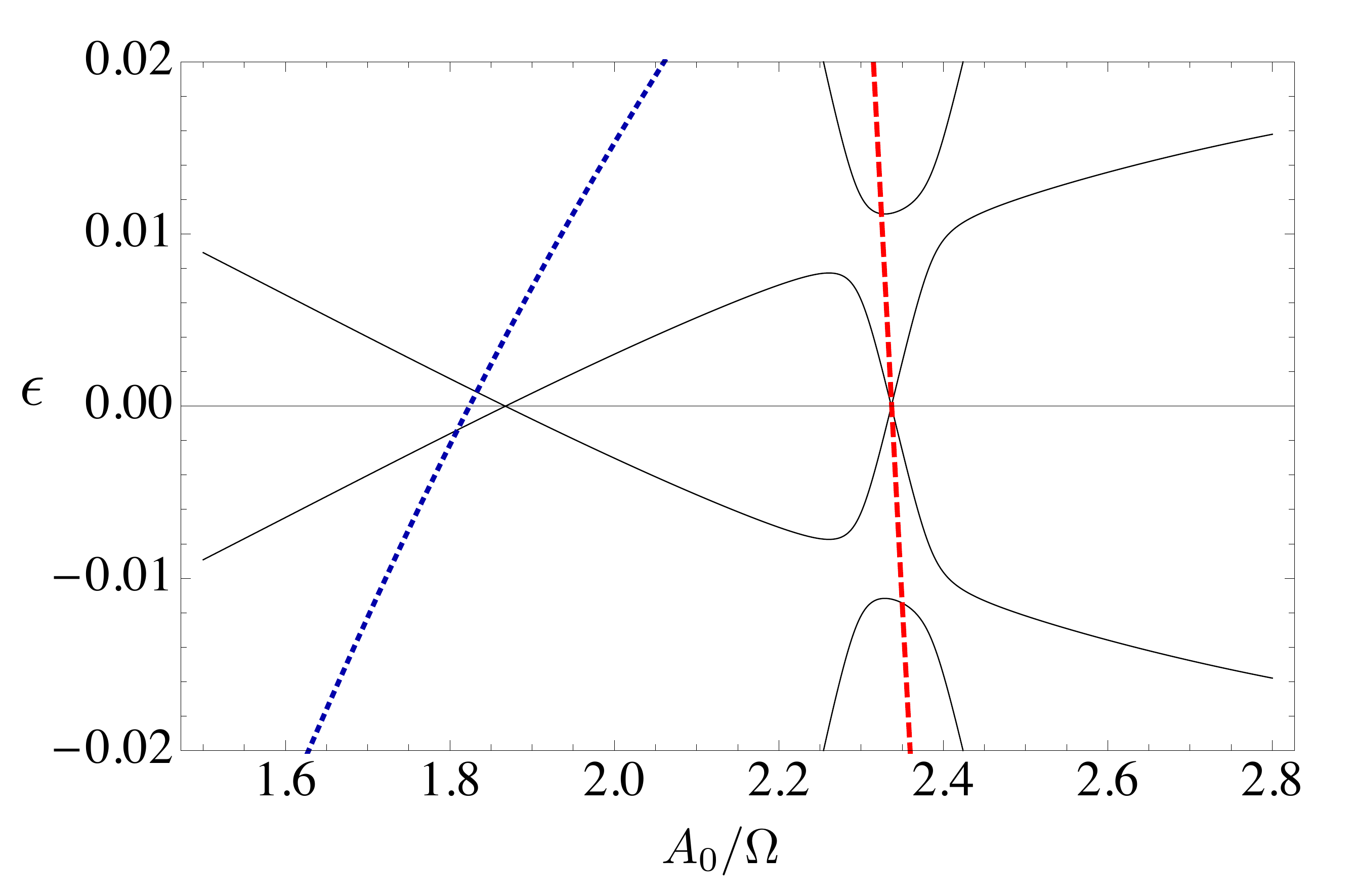}
\protect\caption{\label{fig:splitting-zeros}Lower part of the quasienergy spectrum
for a superconducting triple QD as a function of the amplitude of
the driving. The dotted (blue) line corresponds to $t_{\text{eff}}-\Delta_{\text{eff}}$
and the dashed (red) to $t_{\text{eff}}+\Delta_{\text{eff}}$. We
show that the FMFs appear close to the conditions $\Delta_{\text{eff}}=\pm t_{\text{eff}}$.
Parameters: $\mu_{1}=\mu_{2}=0$, $\mu_{3}=1.5$, $\Delta=1$, $t=0.8$,
$\Omega=4$, $\varphi_{3}-\varphi_{1}=0$. All the energies are in units of $\Delta$, which is set to 1.}
\end{figure}
 Moreover,  the chemical potentials $\mu_{2}$ and $\mu_{3}$ are shifted, such that 
\begin{eqnarray}
{\mu_{2}}_{\text{eff}}&=& 2\frac{\mu_{3}\left(\Delta^{2}-t^{2}\right)}{\Omega^{2}}\sum_{m=1}^{\infty}\frac{{\cal J}_{m}\left(\frac{A_{0}}{\Omega}\right)^{2}}{m^{2}}\ ,\label{eq:mu2eff}\\
{\mu_{3}}_{\text{eff}}&=&\mu_3-2\frac{\mu_{3}\left(\Delta^{2}+t^{2}\right)}{\Omega^{2}}\sum_{m=1}^{\infty}\frac{{\cal J}_{m}\left(\frac{A_{0}}{\Omega}\right)^{2}}{m^{2}}\ .\label{eq:mu3eff}
\end{eqnarray}
The shift in the chemical potentials only changes the localization
of the states (see eq.\ref{eq:states3}) and the effect of the long-range tunneling is small.
In order to probe this, we plot in Fig.\ref{fig:splitting-zeros}
the quasienergy spectrum around zero and the functions $\Delta_{\text{eff}}\pm t_{\text{eff}}$
as a function of the driving amplitude. The sweet spots are very close
to the zeros of these functions, indicating that the effect of $\tau_{1,3}$ is small. 
 Finally, we calculate the localization of the FMFs found in this
configuration. We choose the FMF that appears when $t_{\text{eff}}=\Delta_{\text{eff}}$
(left zero in Fig.\ref{fig:splitting-zeros}). The Majorana pairs
are given by:
\begin{eqnarray}
\gamma_{1} & = & \frac{1}{\sqrt{2}}\left(d_{1}-d_{1}^{\dagger}\right)\ ,\label{eq:FMFs}\\
\gamma_{2} & = & a_{1}\left(d_{1}-d_{1}^{\dagger}\right)+a_{2}\left(d_{2}-d_{2}^{\dagger}\right)+a_{3}\left(d_{3}-d_{3}^{\dagger}\right)\ ,\nonumber 
\end{eqnarray}
with normalization $\sum_{i=1}^{3}2a_{i}^{2}=1$. In the bar
diagram in Fig.\ref{fig:loc} the value of the constants $a_{i}$
for different values of the chemical potential $\mu_{3}$ is plotted.  Interestingly,  Fig.\ref{fig:loc} shows that
for certain values of the chemical potential $\mu_{3}$  the two FMFs are spatially separated and that it is possible to tune
 the position of $\gamma_{2}$. 

\begin{figure}
\centering
\includegraphics[width=1\columnwidth]{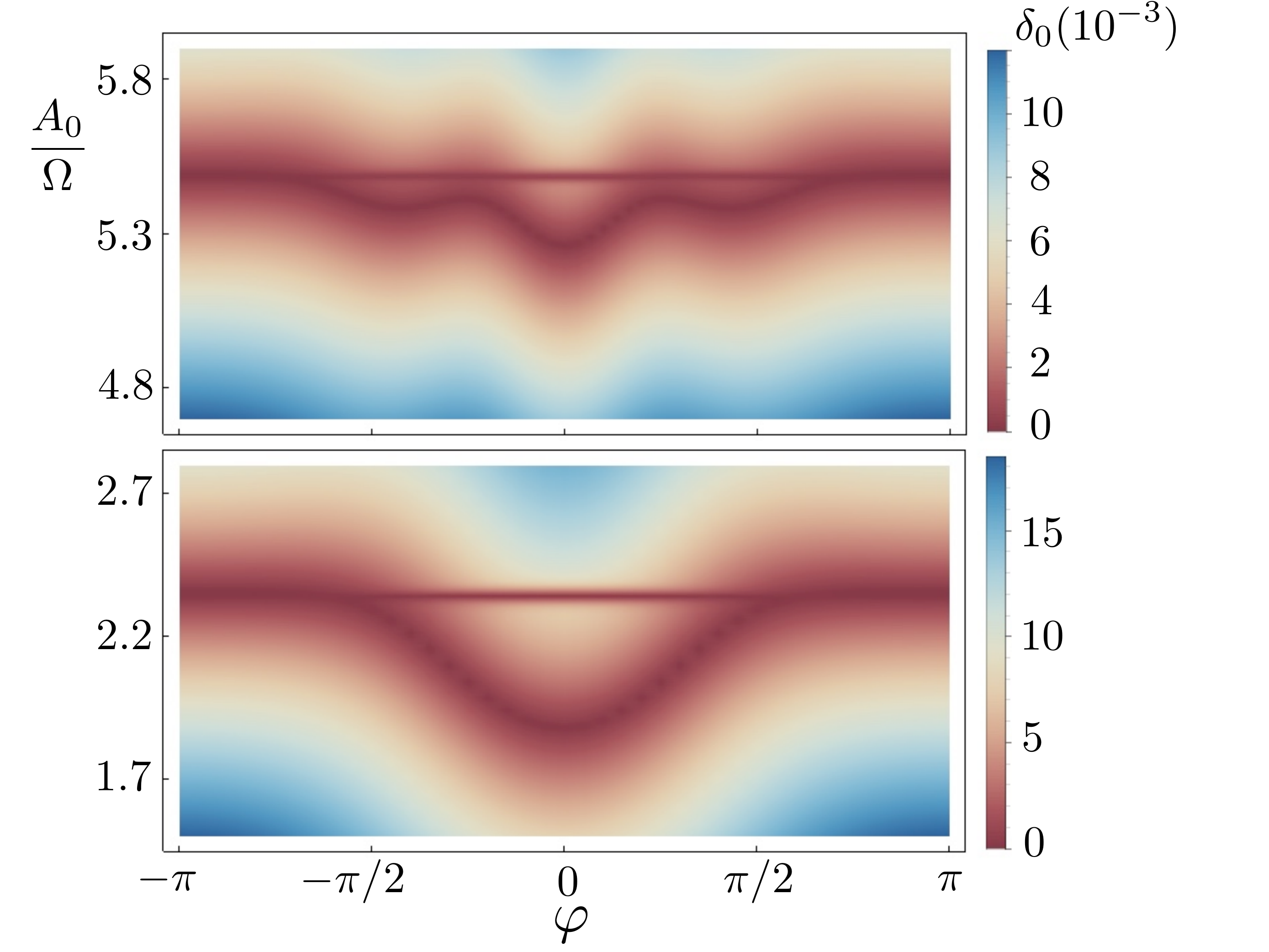}

\protect\caption{\label{fig:splitting-theta}Quasienergy gap $\delta_0$ for a superconducting
triple QD as a function of the amplitude and the relative phase of
the driving fields $\varphi=\varphi_{3}-\varphi_{1}$. The dark regions
correspond to closed gap, it means, zero quasienergy. The plot shows
that one of the sweet spots varies with the phase, while the other
does not. The bottom plot shows the region around the first zero of
the function ${\cal J}_{0}\left(\frac{A_{0}}{\Omega}\right)$ and
the upper plot around the second zero. Parameters: $\mu_{1}=\mu_{2}=0$,
$\mu_{3}=1.5$, $\Delta=1$, $t=0.8$, $\Omega=4$. All the energies are in units of $\Delta$, which is set to 1.}
\end{figure}

Furthermore,  as we will see below, the phase difference between the local ac gate voltages within each dot plays an important role.
Then,  in order to conclude the analysis about the generation of
FMFs in a triple QD configuration, we will show  that the existence of sweet spots 
depends on the relative phase between the driving fields. When the
two fields have opposite phases, $\varphi_{3}-\varphi_{1}=\pi$, the
zero order term of the expansion (\ref{eq:corrections}) does not
change respect to the previous case where  $\varphi_{3}-\varphi_{1}=0$. However, the following
corrections depend on the phase difference. We have calculated the
effective tunneling amplitudes in the case $\varphi_{3}-\varphi_{1}=\pi$
and the result is 
\begin{eqnarray}
t_{\text{eff}} & = & t{\cal J}_{0}\left(\frac{A_{0}}{\Omega}\right)-\frac{2t\left(t^{2}+\Delta^{2}\right)}{\Omega^{2}}f\left(\frac{A_{0}}{\Omega}\right)\ ,\label{eq:teff3dots-pi}\\
\Delta_{\text{eff}} & = & \Delta{\cal J}_{0}\left(\frac{A_{0}}{\Omega}\right)-\frac{2\Delta\left(t^{2}+\Delta^{2}\right)}{\Omega^{2}}f\left(\frac{A_{0}}{\Omega}\right)\ .\label{eq:deltaeff3dots-pi}
\end{eqnarray}
Therefore the functions $\Delta_{\text{eff}}\pm t_{\text{eff}}$ become
zero for the same value of $A_{0}$ in contrast with the previous case for $\varphi=0$. In Fig.\ref{fig:splitting-theta}
we show the gap of the quasienergies as a function of the phase difference
and the amplitude of the driving field. The measurement of this $\varphi$-dependence
would be an important signature of the existence of FMFs.

The existence of these exotic dynamical quasiparticles can be detected
by connecting two metallic leads and measuring transport \cite{Leijnse2012,Deng2015,Flensberg2010,Deng2014}.
The signatures of FMFs will be present in the differential conductance
measurement by the fulfillment of the Floquet sum rule \cite{Kundu2013}.  It is expected that FMFs could be measured by transport by tuning the parameters of the ac driving and therefore the normal and superconducting couplings.

\section{Conclusions}\label{conclu}

To summarize, we have discussed the existence of FMFs in two different
configurations of QDs driven by ac gate voltages and coupled through
superconductor leads. The simplicity of these systems and their tunability in comparison with other proposed setups which provide MFs deserve to consider them as 
suitable solid state devices to host MFs. We have shown
the existence of FMFs by means of  the expansion of an effective Floquet hamiltonian in power series. By modifying the frequency
of the driving field applied to a double QD it is possible to control
the existence of a series of sweet spots. 
Moreover, we analyze as well a  driven triple QD and we  predict the existence of sweet spots as a function of the relative phase of the local drivings. This method for
FMFs generation can be extended to chains of QDs with more than three atoms. One would expect that as the number of QDs increases, the localization of the FMFs changes and eq.(\ref{eq:FMFs}) would be generalized.  Experimentally,  the recent achievements in the fabrication and control of triple \cite{Rogge2009,Granger2010,Busl2013,Sanchez2014} and even quadruple semiconductor QDs \cite{Takakura2014}, also for driven configurations \cite{Braakman2013},  open the avenue for the experimental realization of hybrid configurations with superconductor contacts where  FMFs can be experimentally investigated.

\section*{Acknowledgements}

This volume is devoted to the memory of our friend and colleague Markus Buttiker. One of us, Gloria Platero,  had the big pleasure to meet Markus  more than twenty years ago.
 Markus  was not only a a scientific reference for me (G.P.) but also a good friend, with whom I shared very nice time in workshops and conferences everywhere. I enjoyed many enlighting discussions with him and his wise advices. 
Markus was always very helpful and supported my  research group in Madrid for many years.
 Markus left us. It is a big lost for our scientific community, we will miss him.\\

We acknowledge the Spanish Ministry
of Economy and Competitiveness through project
no. MAT2014-58241-P and the associated FPI scholarship
(M.B.). M. B. thanks the theory division of Max-Planck-Institute for Quantum Optics, where part of this work was realized. 

\section*{References}

\bibliographystyle{elsarticle-num}


\end{document}